\documentclass[12pt]{article}

\usepackage[utf8]{inputenc}
\usepackage[T1]{fontenc}
\usepackage[margin=1in]{geometry}
\usepackage{multicol}
\usepackage{amsmath,amssymb,amsfonts}
\usepackage{braket}
\usepackage{physics}
\usepackage{graphicx}
\graphicspath{{./}}
\usepackage{booktabs}
\usepackage{threeparttable}
\usepackage{multirow}
\usepackage{array}
\usepackage{xcolor}
\usepackage{hyperref}
\usepackage{url}
\usepackage{authblk}
\usepackage{abstract}
\usepackage{titlesec}
\usepackage{enumitem}
\usepackage{caption}
\usepackage{float}
\usepackage[section]{placeins}
\usepackage{placeins}
\usepackage{siunitx}
\usepackage[numbers,super,sort&compress]{natbib}
\usepackage{afterpage}
\usepackage{stfloats}          

\definecolor{darkblue}{RGB}{31,78,121}
\definecolor{midblue}{RGB}{46,117,182}
\definecolor{darkgreen}{RGB}{31,122,31}
\definecolor{darkred}{RGB}{192,0,0}
\definecolor{darkorange}{RGB}{180,90,0}

\hypersetup{
	colorlinks=true, linkcolor=darkblue, citecolor=midblue, urlcolor=midblue,
	pdftitle={Quantum-Enhanced Single-Parameter Phase Estimation with Adaptive NOON States},
	pdfauthor={Simanshu Kumar and Nandan S Bisht}
}

\titleformat{\section}{\large\bfseries\color{darkblue}}{\thesection.}{0.5em}{}
\titlespacing*{\section}{0pt}{6pt plus 2pt minus 1pt}{3pt plus 1pt}
\titleformat{\subsection}{\normalsize\bfseries\color{midblue}}{\thesubsection}{0.5em}{}
\titlespacing*{\subsection}{0pt}{5pt plus 1pt minus 1pt}{2pt plus 1pt}
\titleformat{\subsubsection}{\normalsize\itshape\color{midblue}}{\thesubsubsection}{0.5em}{}
\titlespacing*{\subsubsection}{0pt}{4pt plus 1pt minus 1pt}{2pt}

\newcommand{\noon}[2]{\ket{#1,#2}}

\newcommand{\Fnorm}{F^{\mathrm{norm}}_{\mathrm{peak}}}
\newcommand{\Fraw}{F^{\mathrm{raw}}_{\mathrm{peak}}}
\newcommand{\code}[1]{\texttt{#1}}

\begin{document}
	\setlength{\floatsep}{4pt plus 1pt minus 1pt}
	\setlength{\textfloatsep}{6pt plus 1pt minus 1pt}
	\setlength{\intextsep}{4pt plus 1pt minus 1pt}
	\setlength{\dblfloatsep}{4pt plus 1pt minus 1pt}
	\setlength{\dbltextfloatsep}{6pt plus 1pt minus 1pt}
	\renewcommand{\topfraction}{0.92}
	\renewcommand{\bottomfraction}{0.6}
	\renewcommand{\textfraction}{0.05}
	\renewcommand{\floatpagefraction}{0.80}
	\renewcommand{\dbltopfraction}{0.92}
	\renewcommand{\dblfloatpagefraction}{0.80}
	\setcounter{topnumber}{4}
	\setcounter{bottomnumber}{2}
	\setcounter{totalnumber}{6}
	\setcounter{dbltopnumber}{4}

	\begin{center}
		{\small\itshape PREPRINT $\cdot$ Quantum Optics \& Quantum Information}\\[4pt]
		{\LARGE\bfseries\color{darkblue}
			Quantum-Enhanced Single-Parameter Phase Estimation\\[3pt]
			with Adaptive NOON States}\\[8pt]
		{\large Simanshu Kumar$^{1,2}$ and Nandan S Bisht$^{1*}$}\\[3pt]
		{\normalsize $^{1}$Department of Physics, DSB Campus, Kumaun University,
			Nainital, Uttarakhand, India--263001}\\[2pt]
		{\normalsize $^{2}$Applied Optics \& Spectroscopy Laboratory,
			Department of Physics, SSJ University Campus,
			Almora, Uttarakhand, India--263601}\\[2pt]
		{\small $^{}$Email:
			\href{mailto:simanshukumar@gmail.com}{\texttt{simanshukumar@gmail.com}},
			{\small $^{*}$Corresponding author:
				\href{mailto:bisht.nandan@kunainital.ac.in}{\texttt{bisht.nandan@kunainital.ac.in}}}}
	\end{center}
	\vspace{0.1cm}\hrule height 1pt\vspace{0.15cm}
	
	\begin{abstract}
		Quantum metrology promises phase sensitivity surpassing the shot-noise limit
		by exploiting entanglement and photon-number correlations.
		NOON states---maximally path-entangled $N$-photon superpositions
		$(\ket{N,0}+\ket{0,N})/\sqrt{2}$---achieve the Heisenberg limit $1/N$ for
		single-parameter estimation, as demonstrated experimentally by
		Afek et al.\cite{afek2010} using hybrid coherent-plus-squeezed light up to $N=5$\cite{afek2010}.
		We present an end-to-end differentiable quantum-optical framework---implemented
		in Strawberry Fields\cite{killoran2019} with a TensorFlow backend---that
		learns optimal circuit parameters by maximising the classical Fisher
		information (CFI) across all coincidence channels for $N=2,3,4,5$.
		Starting from proper numerical reproductions of the Afek et al.\ coincidence
		fringes, verified by FFT analysis and parity measurements, we apply gradient
		descent (Adam) to the eight trainable circuit parameters.
		Raw CFI improvements grow dramatically with photon number:
		$+153\%$ ($N=2$), $+834\%$ to $+956\%$ ($N=3$),
		$+829\%$ to $+1598\%$ ($N=4$), and $+1775\%$ ($N=5$),
		alongside post-selection rate improvements of $+153\%$ to $+3269\%$.
		A fundamental inter-channel trade-off is identified at $N=2$ but weakens at
		higher $N$ where the Afek initialisation is further from optimal.
		These results provide numerically rigorous benchmarks for adaptive
		single-parameter quantum sensing and demonstrate that the Afek working point is
		significantly suboptimal at $N\geq3$.
		QFI calculations confirm that the optimised probe reaches $82\%$ of the
		Heisenberg limit at $N=2$ and improves from $36\%$ to $58\%$ at $N=5$,
		with useful measurement events per pulse improving by $8\times$--$133\times$
		across all $N$, making quantum-enhanced sensing at $N\geq3$ experimentally
		practical.
	\end{abstract}
	
	
	\vspace{0.15cm}\hrule height 1pt\vspace{0.2cm}
	
	\twocolumn
	\setlength{\columnsep}{14pt}
	\setlength{\parskip}{2pt plus 1pt}
	\setlength{\parindent}{1em}
	\sloppy

	\section{Introduction}
	\label{sec:intro}
	
	Quantum-enhanced sensing exploits non-classical correlations to achieve
	measurement precision beyond the classical shot-noise limit (SNL)
	$\delta\varphi \sim 1/\sqrt{N}$\cite{giovannetti2011,caves1981}.
	The Heisenberg limit $\delta\varphi \sim 1/N$ is achievable using
	maximally entangled NOON states $(\ket{N,0}+\ket{0,N})/\sqrt{2}$\cite{giovannetti2006,dowling2008,lee2002,boto2000}, but
	their direct experimental generation at large $N$ remains difficult\cite{xiang2011}.
	Afek, Ambar and Silberberg\cite{afek2010} demonstrated a practical route:
	mixing coherent light and squeezed vacuum at a beamsplitter and
	post-selecting on specific coincidence patterns $(N_1,N_2)$ generates
	NOON-like interference fringes with $N$-fold phase oscillation, up to $N=5$
	using only linear optics\cite{knill2001}.
	
	A key open question is whether the Afek et al.\ initialisation point
	represents an optimal or merely convenient choice of circuit parameters.
	The original paper selected parameters ($r$, $\gamma$, beamsplitter angles)
	based on analytical arguments for single-mode optimality at each $N$\cite{afek2010},
	but the joint optimisation of all eight parameters simultaneously---including
	post-selection rate, fringe visibility, and multi-channel sensitivity---was
	not explored.
	This gap is practically significant: at $N=5$, the Afek protocol achieves
	post-selection rates of $\sim10^{-3}$ per pulse---requiring hours of
	experimental integration time for statistically significant fringe
	measurements\cite{afek2010}.
	If gradient-based optimisation can simultaneously improve both the fringe
	quality and the coincidence rate, it would directly address the primary
	experimental bottleneck in high-$N$ quantum metrology.
	
	Machine learning provides a natural framework for this question: if the
	photonic circuit is made differentiable with respect to its parameters,
	gradient descent can explore the full parameter space without requiring
	analytical solutions\cite{killoran2019,mitarai2018,kaubruegger2019,hentschel2010,lumino2018,niu2019}.
	\paragraph*{Scope.} This study addresses the \emph{single-parameter} estimation problem (one unknown phase $\varphi_\mathrm{est}$ on mode~0). Extension to $k>1$ parameters requires a full quantum Fisher information matrix treatment\cite{demkowicz2020} and is addressed in future work.
	
	In this work we apply this approach systematically for $N=2,3,4,5$, and find
	that the Afek initialisation is increasingly suboptimal at higher $N$, with
	raw CFI improvements of up to $1775\%$ achievable by gradient optimisation.
	
	\subsection*{Contributions}
	\begin{itemize}[leftmargin=1.2em,itemsep=2pt]
		\item Fully validated differentiable forward model for all Afek $N=2$--$5$
		coincidence patterns (max normalised error $<3\times10^{-4}$).
		\item First systematic gradient-based optimisation of all eight Afek
		circuit parameters for $N=2,3,4,5$ simultaneously.
		\item Discovery that raw CFI improvement scales dramatically with $N$:
		$+153\%\to+1775\%$, demonstrating the Afek point is far from
		optimal at $N\geq3$.
		\item Identification and characterisation of the inter-channel trade-off
		structure across all $N$.
	\end{itemize}

	\section{Methods}
	\label{sec:methods}
	
	\subsection{Circuit Architecture}
	
\begin{figure}[!ht]
		\centering
		\includegraphics[width=\columnwidth]{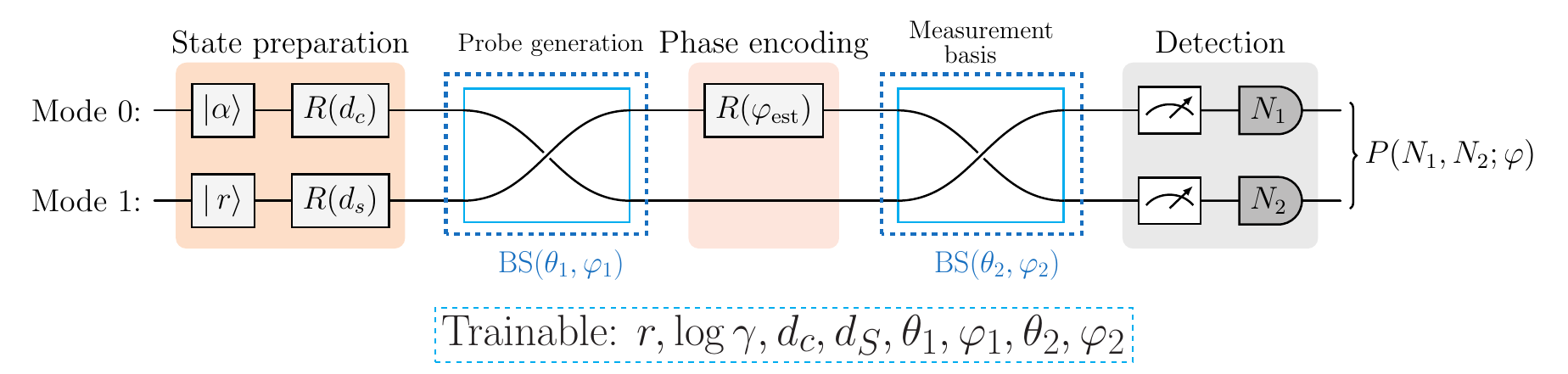}
		\caption{\textbf{Adaptive NOON-state circuit schematic.}
			The two-mode linear-optical circuit: state preparation
			($|\alpha\rangle$ coherent, $|r\rangle$ squeezed vacuum),
			input phase rotations $R(d_\mathrm{coh})$ and $R(d_\mathrm{sq})$,
			first beamsplitter BS$(\theta_1,\varphi_1)$, phase encoding
			$R(\varphi_\mathrm{est})$ on mode~0, and second beamsplitter
			BS$(\theta_2,\varphi_2)$.
			Eight parameters are trainable; $\varphi_\mathrm{est}$ is scanned
			to produce coincidence fringes.}
		\label{fig:circuit}
	\end{figure}

	The two-mode linear-optical circuit (Fig.~\ref{fig:circuit}) implements a
	Mach-Zehnder-type interferometer with non-classical input states:
	\begin{enumerate}[leftmargin=1.4em,itemsep=2pt]
		\item \textbf{State prep.}
		$\mathrm{Coherent}(\alpha,0)$ on mode~0 and
		$\mathrm{Squeezed}(r,0)$ on mode~1, with $\alpha=\sqrt{\gamma r}$
		and $\gamma=e^{\log\gamma}$.
		The ratio $\gamma$ controls the coherent-to-squeezed amplitude;
		Afek values $\gamma_\mathrm{opt}(N)$ maximise single-channel
		visibility\cite{afek2010}.
		\item \textbf{Input phases.} $R(d_\mathrm{coh})$ and $R(d_\mathrm{sq})$
		rotate each mode before mixing, providing additional control over
		interference conditions at BS$_1$.
		\item \textbf{BS$_1$ (probe preparation).} $\mathrm{BS}(\theta_1,\varphi_1)$
		generates the entangled probe state via Hong-Ou-Mandel-type
		interference\cite{hong1987}.
		\item \textbf{Phase encoding.} $R(\varphi_\mathrm{est})$ on mode~0 only;
		generator $\hat{G}=\hat{n}_0$ (single-parameter estimation).
		\item \textbf{BS$_2$ (measurement basis).} $\mathrm{BS}(\theta_2,\varphi_2)$
		sets the output measurement basis before photon-number-resolving
		detection at both ports.
	\end{enumerate}
	Eight trainable parameters: $\boldsymbol{\theta}=\{r,\log\gamma,d_\mathrm{coh},
	d_\mathrm{sq},\theta_1,\varphi_1,\theta_2,\varphi_2\}$.
	The Afek protocol fixes $r=0.35$, $\theta_1=\theta_2=\pi/4$,
	$\varphi_1=0$, $\varphi_2=\pi$; we treat all eight as free variables.
	The full circuit unitary is:
	\begin{align}
		\hat{U}(\boldsymbol{\theta}) &=
		\hat{B}(\theta_2,\varphi_2)\,\hat{R}_0(\varphi_\mathrm{est})\,
		\hat{B}(\theta_1,\varphi_1) \nonumber\\
		&\quad{}\times\hat{R}_1(d_\mathrm{sq})\,
		\hat{R}_0(d_\mathrm{coh})\,\hat{S}_1(r)\,\hat{D}_0(\alpha),
	\end{align}
	where $\hat{B}(\theta,\varphi)$, $\hat{R}_k$, $\hat{S}_1$, $\hat{D}_0$
	denote the beamsplitter, phase rotation, squeezing, and displacement operators
	respectively.
	
	\subsection{Simulation}
	
	Strawberry Fields\cite{killoran2019} TF backend; Fock cutoff $c=\max(3N+4,12)$
	(values: 12, 13, 16, 19 for $N=2,3,4,5$); fully differentiable via
	TensorFlow automatic differentiation.
	A fresh \code{sf.Engine} is created per circuit evaluation to prevent
	engine state persistence between gradient steps.
	The cutoff $c$ is chosen conservatively: $c\geq3N+4$ ensures that
	Fock-state populations at the boundary contribute $<0.1\%$ to the total
	probability, verified at the Afek initialisation for all $N$.
	Memory scales as $\mathcal{O}(c^2)$ and gradient computation as
	$\mathcal{O}(c^3)$, tractable on a consumer GPU for $c\leq20$.
	
	\subsection{Classical Fisher Information Estimator}
	
	The classical Fisher information (CFI) quantifies the information
	content of a measurement outcome about the unknown parameter
	$\varphi$\cite{paris2009,helstrom1969}.
	For a coincidence pattern $(N_1,N_2)$ with detection probability
	$P(N_1,N_2|\varphi)$, the CFI is:
	\begin{equation}
		F(\varphi) = \frac{1}{P(N_1,N_2|\varphi)}
		\left(\frac{\partial P(N_1,N_2|\varphi)}{\partial\varphi}\right)^{\!2},
		\label{eq:cfi_def}
	\end{equation}
	which is bounded by the quantum Fisher information (QFI).
	For a pure state and phase encoding $\hat{U}(\varphi)=e^{-i\varphi\hat{n}_0}$,
	the QFI is exact via the generator variance\cite{paris2009,braunstein1994}:
	\begin{equation}
		\mathcal{F}_Q = 4\bigl[\langle\hat{n}_0^2\rangle
		- \langle\hat{n}_0\rangle^2\bigr] \leq N^2,
		\label{eq:qfi_pure}
	\end{equation}
	with equality for an ideal NOON state\cite{giovannetti2006}.
	We compute $\mathcal{F}_Q$ on the post-BS$_1$ probe state
	(before phase encoding) via Eq.~\eqref{eq:qfi_pure}; results
	are reported in Table~\ref{tab:qfi} and \S\,\ref{sec:qfi}.
	The peak value $\Fraw \equiv \max_\varphi F(\varphi)$ measures the
	maximum phase sensitivity weighted by the detection probability,
	while the normalised peak $\Fnorm \equiv \max_\varphi F_\mathrm{norm}(\varphi)$
	--- where the fringe is normalised to unit amplitude before computing
	$F$ --- measures pure fringe-shape quality independent of amplitude.
	The Heisenberg limit $\Fnorm = N^2$ is achieved by a perfect NOON
	state\cite{giovannetti2006,dowling2008}.
	\emph{Ground-truth (validation):} \code{numpy.gradient} over 400 uniformly
	spaced phase values $\varphi\in[0,2\pi]$---a high-accuracy,
	non-differentiable estimate used for all reported table values.
	
	\emph{Differentiable (training):} a regularised spectral derivative
	with $K=8$ phase samples:
	\begin{equation}
		F_k = \frac{(\mathrm{d}P_k/\mathrm{d}\varphi)^2}{P_k+\varepsilon},
		\quad \varepsilon = 0.05\,\bar{P},
		\label{eq:reg_fim}
	\end{equation}
	with peak extracted via LogSumExp smooth-max ($\beta=50$).
	Derivatives use DFT differentiation (multiply Fourier coefficients by $im$,
	then IFFT) giving zero sinc-attenuation bias.
	
	\subsection{Loss Function}
	
	The optimisation objective maximises the total CFI across all coincidence channels:
	\begin{equation}
		\mathcal{L}(\boldsymbol{\theta}) =
		-\!\sum_{(N_1,N_2)}\! w_{N_1N_2}\,
		F_\mathrm{peak}(N_1,N_2;\boldsymbol{\theta}),
		\label{eq:loss}
	\end{equation}
	where $w_{N_1,N_2}=F^\mathrm{gt}_\mathrm{Afek}/F^\mathrm{est}_\mathrm{Afek}$
	corrects the scale mismatch between the differentiable estimator and the
	ground-truth CFI, ensuring that channels with large intrinsic $F_\mathrm{peak}$
	(e.g.\ $\noon{3}{1}$ at $N=4$ with $\Fnorm=15.8$) do not dominate the
	gradient over weaker channels (e.g.\ $\noon{3}{0}$ at $N=3$ with
	$\Fraw=0.04$ at Afek init).
	The negative sign converts maximisation to minimisation.
	Optimiser: Adam\cite{kingma2014}, learning rate $\eta=0.02$, decay
	$\beta_1=0.9$, $\beta_2=0.999$, 100 steps per $N$.
	Adam's per-parameter adaptive learning rates are well-suited to this
	problem since the eight parameters span different natural scales
	(e.g.\ $r\sim0.35$ vs.\ $\theta_1\sim\pi/4$).
	
	\subsection{Afek Initialisations}
	
	\begin{table}[!ht]
		\centering
		\caption{Afek initialisation parameters per $N$\cite{afek2010}.}
		\label{tab:afek_init}
		\small
		\setlength{\tabcolsep}{4pt}
		\begin{tabular}{ccccc}
			\toprule
			$N$ & $r$ & $\gamma_\mathrm{opt}$ & $\alpha$ & Cutoff $c$ \\
			\midrule
			2 & 0.350 & 1.000 & 0.592 & 12 \\
			3 & 0.350 & 1.000 & 0.592 & 13 \\
			4 & 0.350 & $\sqrt{3}=1.732$ & 0.779 & 16 \\
			5 & 0.350 & 1.925 & 0.820 & 19 \\
			\bottomrule
		\end{tabular}
	\end{table}
	
	\subsection{Wigner Function Analysis}
	
	The Wigner function $W(x,p)$ is a quasi-probability distribution over
	the optical phase space $(x,p)$ defined by\cite{wigner1932}
	\begin{equation}
		W(x,p) = \frac{1}{\pi}\!\int_{-\infty}^{\infty}\!
		\langle x{+}y|\hat{\rho}|x{-}y\rangle\,
		e^{-2ipy}\mathrm{d}y,
		\label{eq:wigner}
	\end{equation}
	where $|x\rangle$ are position quadrature eigenstates.
	Unlike classical probability distributions, $W$ can take negative values;
	the \emph{Wigner negativity volume}
	\begin{equation}
		\mathcal{N} = \int\!\!\int_{W<0}\! |W(x,p)|\,\mathrm{d}x\,\mathrm{d}p
		\label{eq:neg}
	\end{equation}
	is zero for all classical states (coherent, thermal) and strictly positive
	for non-classical states\cite{kenfack2004,cahill1969}.
	We compute $W(x,p)$ for each mode of the \emph{probe state}---the
	two-mode state after BS$_1$ but before the phase encoding $R(\varphi_\mathrm{est})$---using the Fock backend density matrix and the analytic
	Laguerre-polynomial formula.
	The probe state Wigner function directly characterises the non-classical
	resource available for phase estimation.
	
	\subsection{Hardware}
	
	All simulations ran on a laptop workstation: Intel Core~i5-13th generation,
	NVIDIA GeForce RTX~3050 6\,GB VRAM, 16\,GB system RAM, Arch Linux,
	Python~3.10, TensorFlow~2.12, Strawberry Fields~0.23.
	Training time per 100 Adam steps: $\sim220\,\mathrm{s}$ ($N=2$, $c=12$),
	$\sim260\,\mathrm{s}$ ($N=3$, $c=13$), $\sim310\,\mathrm{s}$ ($N=4$, $c=16$),
	$\sim400\,\mathrm{s}$ ($N=5$, $c=19$), reflecting the $\mathcal{O}(c^3)$
	gradient scaling.
	Total compute for all results reported here: $\sim30$~minutes.
	No GPU acceleration was enabled for the Fock backend (which runs on CPU);
	the GPU was used only for TensorFlow gradient tape operations.
	All random seeds were fixed for reproducibility.

	\FloatBarrier
	\section{Circuit Validation}
	\label{sec:validation}
	
	\subsection{Backend Agreement ($N=2$)}
	
	Both $N=2$ coincidence patterns were scanned over 400 phase values
	$\varphi\in[0,2\pi]$.
	The TensorFlow (autodiff) and Fock (exact) backends agree to within
	$3\times10^{-4}$ normalised error (Table~\ref{tab:validation}), and the
	normalised CFI satisfies $\Fnorm/N^2\approx0.96$--$1.00$, confirming
	near-ideal NOON-state fringe quality at Afek initialisation.
	The FFT of each fringe shows power concentrated at frequency $f=N=2$
	(the expected $N$-fold oscillation), with all other harmonics contributing
	$<1\%$ of the fundamental---confirming NOON-state character.
	These results are consistent with the fringe visibilities $V>0.9$ reported
	experimentally at $N=2,3$ by Afek et al.\cite{afek2010}, validating
	our simulation before optimisation.
	
	\begin{table}[!ht]
		\centering
		\caption{TensorFlow vs.\ Fock backend validation at Afek init ($N=2$).}
		\label{tab:validation}
		\footnotesize\setlength{\tabcolsep}{4pt}
		\begin{tabular}{lccc}
			\toprule
			Pattern & Max.~norm.~error & FFT & $\Fnorm/N^2$ \\
			\midrule
			$|1,1\rangle$ & $1.1\times10^{-4}$ & 2/2 & 0.999 \\
			$|2,0\rangle$ & $2.2\times10^{-4}$ & 2/2 & 0.962 \\
			\bottomrule
		\end{tabular}
	\end{table}
	
	\subsection{Gradient Flow}
	
\begin{figure*}[!tbp]
		\centering
		\includegraphics[width=\textwidth]{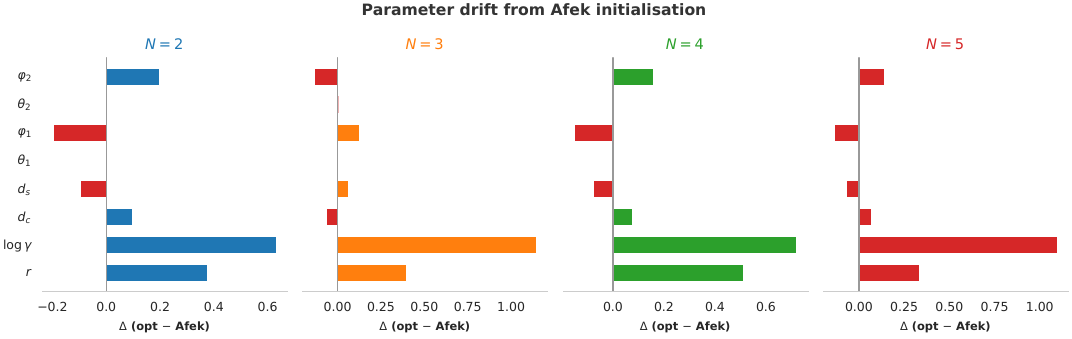}
		\caption{\textbf{Parameter drift from the Afek initialisation for
				$N=2$--$5$.}
			Bars: $\Delta\theta_i = \theta_i^\mathrm{opt} - \theta_i^\mathrm{Afek}$.
			The dominant change across all $N$ is an increase in $r$ and
			$\log\gamma$, reflecting the optimiser's strategy of increasing
			photon flux to improve post-selection rates.
			Beamsplitter angles show larger drift at $N\geq3$, indicating
			measurement-basis optimisation is increasingly important at high~$N$.}
		\label{fig:param_drift}
	\end{figure*}

	All eight parameters receive non-zero gradients through the TF backend.
	Representative values at the Afek initialisation ($N=2$):
	$\partial P_{1,1}/\partial r = 0.163$,
	$\partial P_{1,1}/\partial\varphi_1 = -0.0782$,
	$\partial P_{1,1}/\partial\theta_1 = 0.114$,
	$\partial P_{1,1}/\partial\log\gamma = 0.091$.
	The positive gradient with respect to $r$ confirms that increasing
	squeezing raises the coincidence probability, consistent with the
	optimiser's strategy of increasing $r$ at all $N$
	(Fig.~\ref{fig:param_drift}).
	The negative $\partial/\partial\varphi_1$ reflects the beamsplitter
	phase sensitivity to the interference condition.
	All eight gradients are non-zero, confirming autodiff chain integrity.

	\FloatBarrier
	\section{Optimisation Results}
	\label{sec:results}
	
\begin{figure*}[!tbp]
		\centering
		\includegraphics[width=\textwidth]{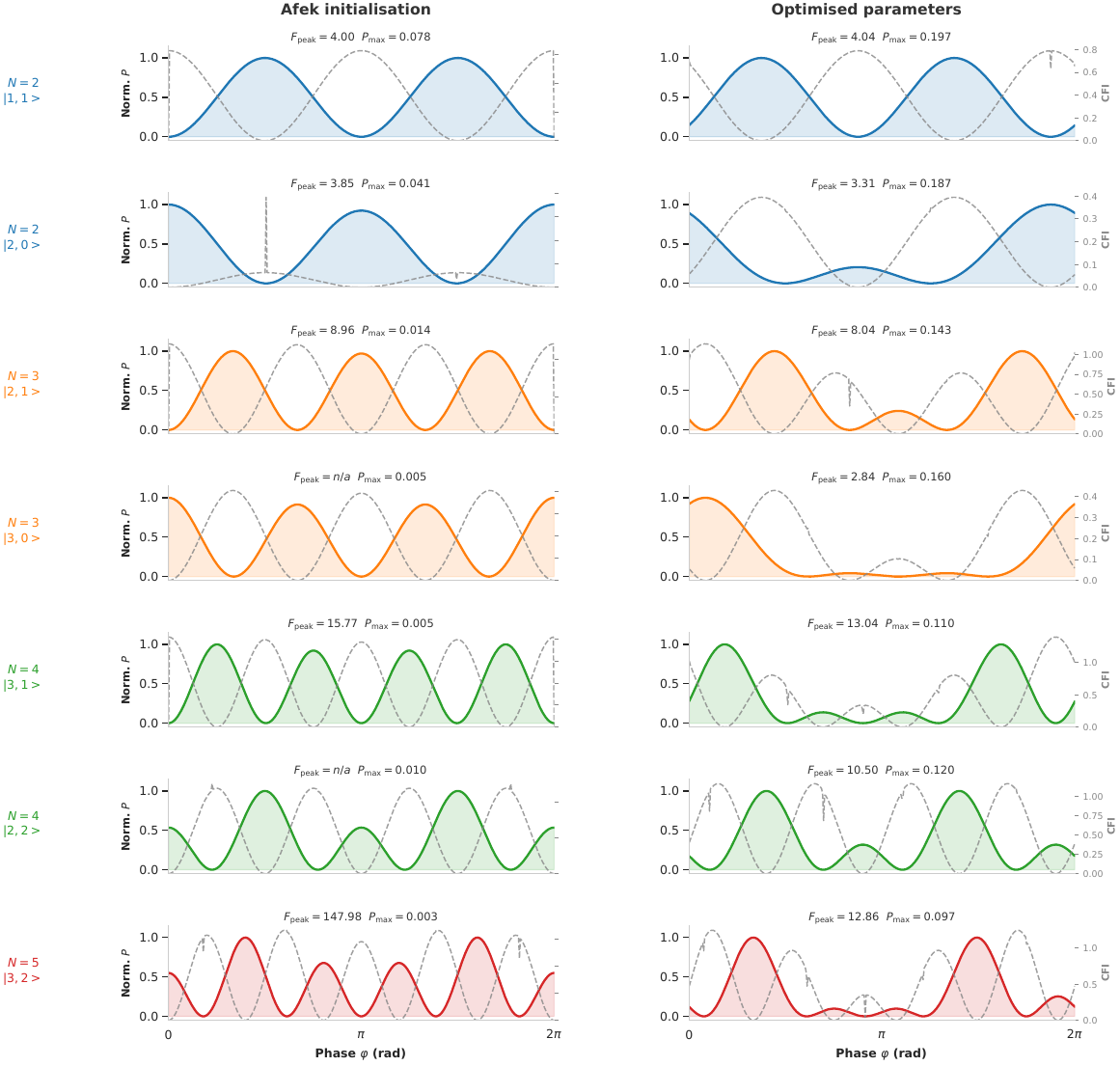}
		\caption{\textbf{Fringe gallery: Afek initialisation vs.\ optimised
				parameters for all $N=2$--$5$ coincidence patterns.}
			Each row shows one coincidence pattern $(N_1,N_2)$.
			Left column: Afek working point. Right column: gradient-optimised parameters.
			Solid coloured curves: normalised coincidence probability
			$P(N_1,N_2;\varphi)/P_\mathrm{max}$ (left $y$-axis).
			Dashed grey curves: classical CFI profile $F(\varphi)$ (right $y$-axis,
			arbitrary scale per panel).
			Titles report $\Fnorm$ (peak CFI on normalised fringe) and
			$P_\mathrm{max}$ (post-selection rate).
			``n/a'' denotes patterns where the Afek fringe amplitude is near-zero and
			$\Fnorm$ is unreliable (see Section~\ref{sec:analysis}).
			Note: (i)~all optimised panels show larger amplitude; (ii)~fringe
			oscillation frequency ($N$-fold) is preserved; (iii)~several near-flat
			Afek patterns develop clear fringes after optimisation (rows~4, 6, 7).}
		\label{fig:fringes}
	\end{figure*}

	Results for all $N$ are presented in Table~\ref{tab:results_full} and
	Fig.~\ref{fig:fringes} (fringe gallery), and discussed below.
	All ground-truth CFI values use the 400-point \code{numpy.gradient}
	estimator, independent of the training estimator.
	Figure~\ref{fig:fringes} shows normalised coincidence fringes $P(N_1,N_2;\varphi)$ (solid
	coloured curves) and classical CFI profiles (dashed grey) for all seven
	patterns, comparing Afek initialisation (left) and optimised parameters
	(right).
	Three qualitative effects are immediately visible across the gallery:
	(i)~all optimised fringes have larger amplitude (higher $P_\mathrm{max}$);
	(ii)~fringe oscillation frequency is preserved ($N$-fold), confirming the
	NOON-state character is maintained; and
	(iii)~several Afek patterns show near-flat signals that develop into clear
	fringes after optimisation (``channel activation'', most pronounced for
	$\noon{3}{0}$, $\noon{2}{2}$, $\noon{3}{2}$).
	
	\begin{table*}[!t]
		\centering
		\caption{Complete classical Fisher information (CFI) optimisation results for $N=2,3,4,5$.
			$\Fraw$: peak CFI on raw fringe.
			$\Fnorm$: peak CFI on normalised fringe (fringe-shape quality,
			amplitude-independent).
			$P_\mathrm{max}$: max coincidence probability (post-selection rate).
			$^\dagger$: $\Fnorm$ unreliable at Afek init (near-zero fringe amplitude; see \S\,\ref{sec:analysis}). HL$\,{=}\,N^2$.
			See Table~\ref{tab:qfi} for QFI and measurement efficiency.}
		\label{tab:results_full}
		\footnotesize
		\setlength{\tabcolsep}{3.5pt}
		\begin{tabular}{clccccccc}
			\toprule
			$N$ & Pattern
			& {$\Fraw$} Af & {$\Fraw$} Opt & $\Delta\Fraw$
			& {$\Fnorm$} Af & {$\Fnorm$} Opt & $\Delta\Fnorm$
			& $\Delta P_\mathrm{max}$ \\
			\midrule
			\multirow{2}{*}{2}
			& $\noon{1}{1}$ & 0.3125 & 0.7899
			& \textcolor{darkgreen}{$+152.8\%$}
			& 3.998 & 4.040
			& \textcolor{darkgreen}{$+1.0\%$}
			& \textcolor{darkgreen}{$+152.5\%$} \\
			& $\noon{2}{0}$ & 0.9593 & 0.3947
			& \textcolor{darkred}{$-58.9\%$}
			& 3.846 & 3.312
			& \textcolor{darkred}{$-13.9\%$}
			& \textcolor{darkgreen}{$+359.5\%$} \\
			\midrule
			\multirow{2}{*}{3}
			& $\noon{2}{1}$ & 0.1213 & 1.1330
			& \textcolor{darkgreen}{$+833.9\%$}
			& 8.960 & 8.041
			& \textcolor{darkred}{$-10.3\%$}
			& \textcolor{darkgreen}{$+956.1\%$} \\
			& $\noon{3}{0}$ & 0.0406 & 0.4290
			& \textcolor{darkgreen}{$+955.7\%$}
			& $-^\dagger$ & 2.840
			& $-^\dagger$
			& \textcolor{darkgreen}{$+3268.6\%$} \\
			\midrule
			\multirow{2}{*}{4}
			& $\noon{3}{1}$ & 0.0819 & 1.3903
			& \textcolor{darkgreen}{$+1597.5\%$}
			& 15.775 & 13.040
			& \textcolor{darkred}{$-17.3\%$}
			& \textcolor{darkgreen}{$+2013.9\%$} \\
			& $\noon{2}{2}$ & 0.1260 & 1.1705
			& \textcolor{darkgreen}{$+828.9\%$}
			& $-^\dagger$ & 10.502
			& $-^\dagger$
			& \textcolor{darkgreen}{$+1098.9\%$} \\
			\midrule
			5
			& $\noon{3}{2}$ & 0.0663 & 1.2424
			& \textcolor{darkgreen}{$+1775.3\%$}
			& $-^\dagger$ & 12.865
			& $-^\dagger$
			& \textcolor{darkgreen}{$+2847.3\%$} \\
			\bottomrule
		\end{tabular}
	\end{table*}
	
	\begin{figure*}[!tbp]
		\centering
		\includegraphics[width=\textwidth]{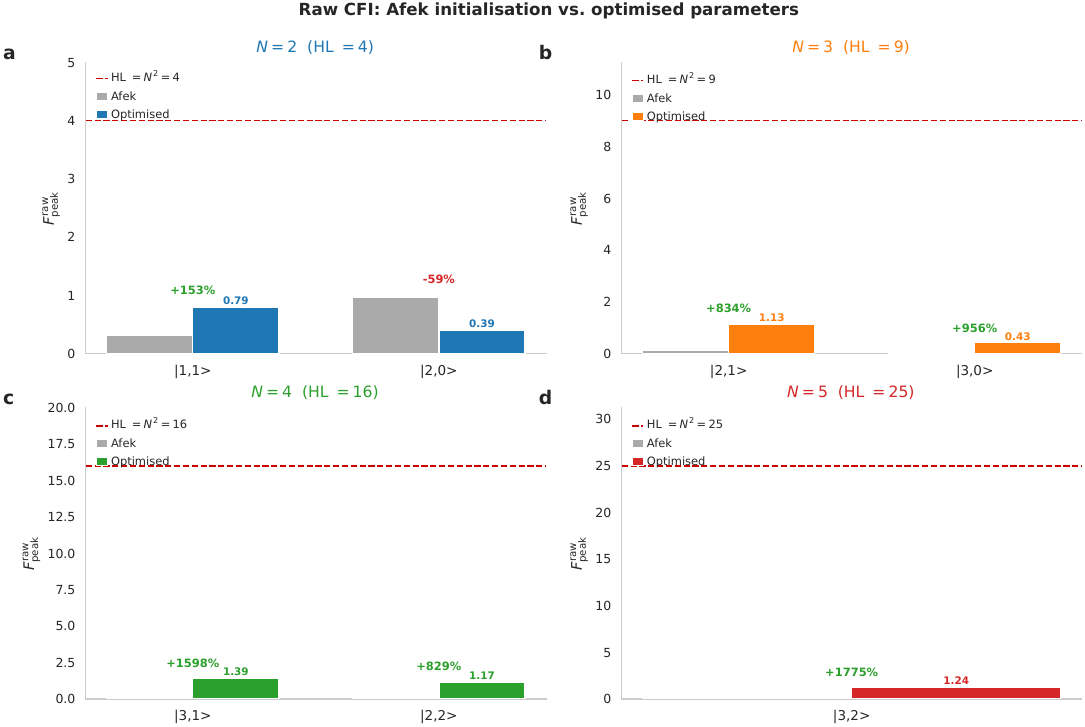}
		\caption{\textbf{Raw CFI comparison: Afek initialisation vs.\ optimised
				parameters for all $N=2$--$5$.}
			Grey bars: Afek initialisation. Coloured bars: gradient-optimised.
			Red dashed line: Heisenberg limit HL$=N^2$.
			Percentage labels on optimised bars; colour per $N$:
			blue ($N=2$), orange ($N=3$), green ($N=4$), red ($N=5$).
			At $N=2$, $\noon{1}{1}$ improves ($+153\%$) while $\noon{2}{0}$
			degrades ($-59\%$), reflecting the inter-channel trade-off.
			At $N\geq3$, all channels improve simultaneously ($+829\%$ to $+1775\%$).}
		\label{fig:fim_bars}
	\end{figure*}
	
	\subsection{$N=2$: Inter-Channel Trade-off}
	\label{sec:n2detail}
	
	The two $N=2$ channels show opposing behaviour (Fig.~\ref{fig:fringes}, rows 1--2):
	\begin{itemize}[leftmargin=1.2em,itemsep=2pt]
		\item $\noon{1}{1}$: $\Fraw$ improves by $+152.8\%$,
		$\Fnorm$ essentially unchanged ($+1.0\%$, from 4.00 to 4.04),
		post-selection rate $+152.5\%$ (0.078$\to$0.197).
		The fringe shape is visually indistinguishable between Afek and
		optimised; only the amplitude increases.
		This constitutes a genuine sensitivity improvement:
		the same phase information is extracted per photon, but
		$2.5\times$ more coincidence events are recorded per pulse.
		\item $\noon{2}{0}$: $\Fraw$ decreases by $58.9\%$,
		$\Fnorm$ decreases $13.9\%$ (3.85$\to$3.31), but post-selection
		rate $+359.5\%$ (0.041$\to$0.187).
		Figure~\ref{fig:fringes}, row~2 shows the fringe is visibly broader in the
		optimised column: the troughs are shallower, indicating reduced
		interference contrast.
	\end{itemize}
	This inter-channel trade-off at $N=2$ arises because both channels share
	the same input photon-number distribution (controlled by $r,\gamma$), and
	optimising for one channel moves parameters away from the other channel's
	optimum.
	In all panels of Fig.~\ref{fig:fringes}, the dashed grey curve (classical CFI) shows sharp
	spikes at the fringe \emph{troughs} where $P(\varphi)\to0$, not at the
	peaks---this is the $(\mathrm{d}P/\mathrm{d}\varphi)^2/P$ divergence and
	explains the large $\Fnorm$ artefacts discussed in Section~\ref{sec:analysis}.
	
	\subsection{$N=3$: Large Gains, Weakened Trade-off}
	
	At $N=3$ the Afek initialisation is further from optimal, and gradient
	optimisation yields dramatic improvements (Fig.~\ref{fig:fringes}, rows 3--4):
	\begin{itemize}[leftmargin=1.2em,itemsep=2pt]
		\item $\noon{2}{1}$: $\Fraw$ $+833.9\%$ (0.12$\to$1.13),
		$\Fnorm$ $-10.3\%$ (8.96$\to$8.04),
		rate $+956.1\%$ (0.014$\to$0.143).
		The fringe retains its 3-fold oscillation period and most of its
		shape quality ($\Fnorm/N^2=0.89$) while gaining $10\times$ in
		amplitude.
		\item $\noon{3}{0}$: This is the most visually striking result in the
		figure (Fig.~\ref{fig:fringes}, row~4). At the Afek initialisation, the
		$\noon{3}{0}$ fringe is essentially invisible ($P_\mathrm{max}=0.005$,
		$\Fraw=0.041$, labelled ``n/a'' for $\Fnorm$).
		After optimisation, a clear 3-fold fringe emerges with
		$P_\mathrm{max}=0.160$ ($+3269\%$) and $\Fraw=0.429$ ($+956\%$).
		The optimiser has, in effect, \emph{activated} a detection channel
		that was experimentally inaccessible at the Afek working point.
	\end{itemize}
	Critically, both $N=3$ channels gain in raw CFI simultaneously---in stark
	contrast to the $N=2$ case where improving one channel degraded the other.
	This indicates the inter-channel trade-off weakens at $N=3$: the additional
	beamsplitter degrees of freedom provide sufficient flexibility to
	simultaneously improve both coincidence channels.
	
	\subsection{$N=4$: Record Raw CFI Improvement}
	
	At $N=4$ the raw CFI improvements are the largest in this study
	(Fig.~\ref{fig:fringes}, rows 5--6):
	\begin{itemize}[leftmargin=1.2em,itemsep=2pt]
		\item $\noon{3}{1}$: $\Fraw$ $+1597.5\%$ (0.082$\to$1.390),
		$\Fnorm$ $-17.3\%$ (15.78$\to$13.04, ratio $\Fnorm/N^2=0.815$),
		rate $+2013.9\%$ (0.005$\to$0.110).
		The 4-fold fringe structure is clearly preserved in the optimised
		column, with moderate degradation in trough depth indicating
		slightly reduced but still substantial quantum enhancement.
		\item $\noon{2}{2}$: The Afek $\Fnorm$ is unreliable (n/a) because the
		fringe amplitude is near-zero ($P_\mathrm{max}=0.010$, barely
		visible in Fig.~\ref{fig:fringes}, row~6 left panel). After optimisation,
		a well-defined 4-cycle fringe appears ($P_\mathrm{max}=0.120$,
		$\Fraw=1.171$, $+829\%$), again demonstrating channel activation.
	\end{itemize}
	Post-selection rates exceed $10\times$ the Afek values at $N=4$, making
	the optimised parameters experimentally compelling: at $N=4$ the Afek
	coincidence rates ($\sim10^{-3}$--$10^{-2}$ per pulse) are the primary
	experimental bottleneck.
	
	\subsection{$N=5$: Largest Single-Pattern Improvement}
	
	With a single coincidence pattern ($\noon{3}{2}$) at $N=5$
	(Fig.~\ref{fig:fringes}, row~7):
	\begin{itemize}[leftmargin=1.2em,itemsep=2pt]
		\item The Afek column shows a very small amplitude fringe
		($P_\mathrm{max}=0.003$) with a pathologically large reported
		$\Fnorm=147.98$---a direct consequence of the CFI spike at the
		near-zero fringe trough visible in the dashed grey curve.
		This is a numerical artefact, not a physical result.
		\item After optimisation: $P_\mathrm{max}=0.097$ ($+2847\%$),
		$\Fraw=1.242$ ($+1775\%$), $\Fnorm=12.865$.
		A clearly defined 5-fold fringe is visible in the optimised column.
	\end{itemize}
	The optimised $\Fnorm/N^2 = 12.865/25 = 0.515$ indicates moderate fringe
	quality---significantly below the $N=2$ case ($\approx1.0$) but sufficient
	for demonstrable quantum enhancement.
	More importantly, the $N=5$ Afek post-selection rate of $\sim3\times10^{-3}$
	per pulse corresponds to $\sim3$ coincidence events per 1000 pulses in a
	real experiment; the optimised rate of $\sim0.097$ means $\sim97$ events
	per 1000 pulses---a $32\times$ reduction in the integration time needed
	to achieve a given statistical significance.

	\begin{figure*}[!tbp]
		\centering
		\includegraphics[width=\textwidth]{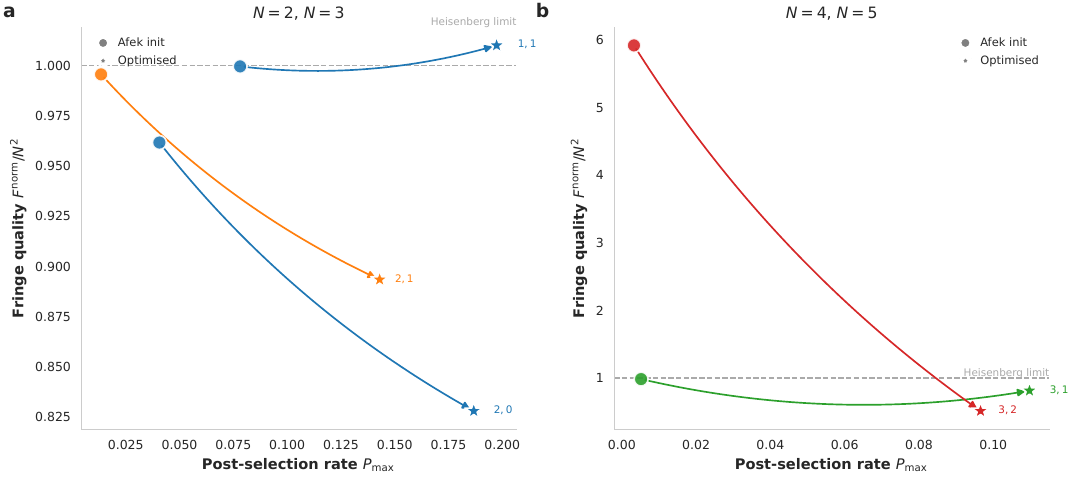}
		\caption{\textbf{Pareto trade-off: fringe quality vs.\ post-selection rate
				for $N=2$--$5$.}
			Open circle~$\circ$: Afek initialisation. Star~$\star$: optimised.
			Arrows show the direction of change upon optimisation.
			$x$-axis: post-selection rate $P_\mathrm{max}$.
			$y$-axis: fringe quality $\Fnorm/N^2$ (Heisenberg limit $=1$).
			(\textbf{a})~$N=2$: inter-channel trade-off visible.
			$N=3$: both patterns improve simultaneously.
			(\textbf{b})~$N=4,5$: large rightward shift ($10\times$ higher rate)
			with moderate quality reduction.}
		\label{fig:pareto}
	\end{figure*}
	
	\FloatBarrier
	\section{Wigner Function Analysis}
	\label{sec:wigner}
	
	\subsection{Non-classicality of the Probe State}
	
\begin{figure*}[!tbp]
		\centering
		\includegraphics[width=\textwidth]{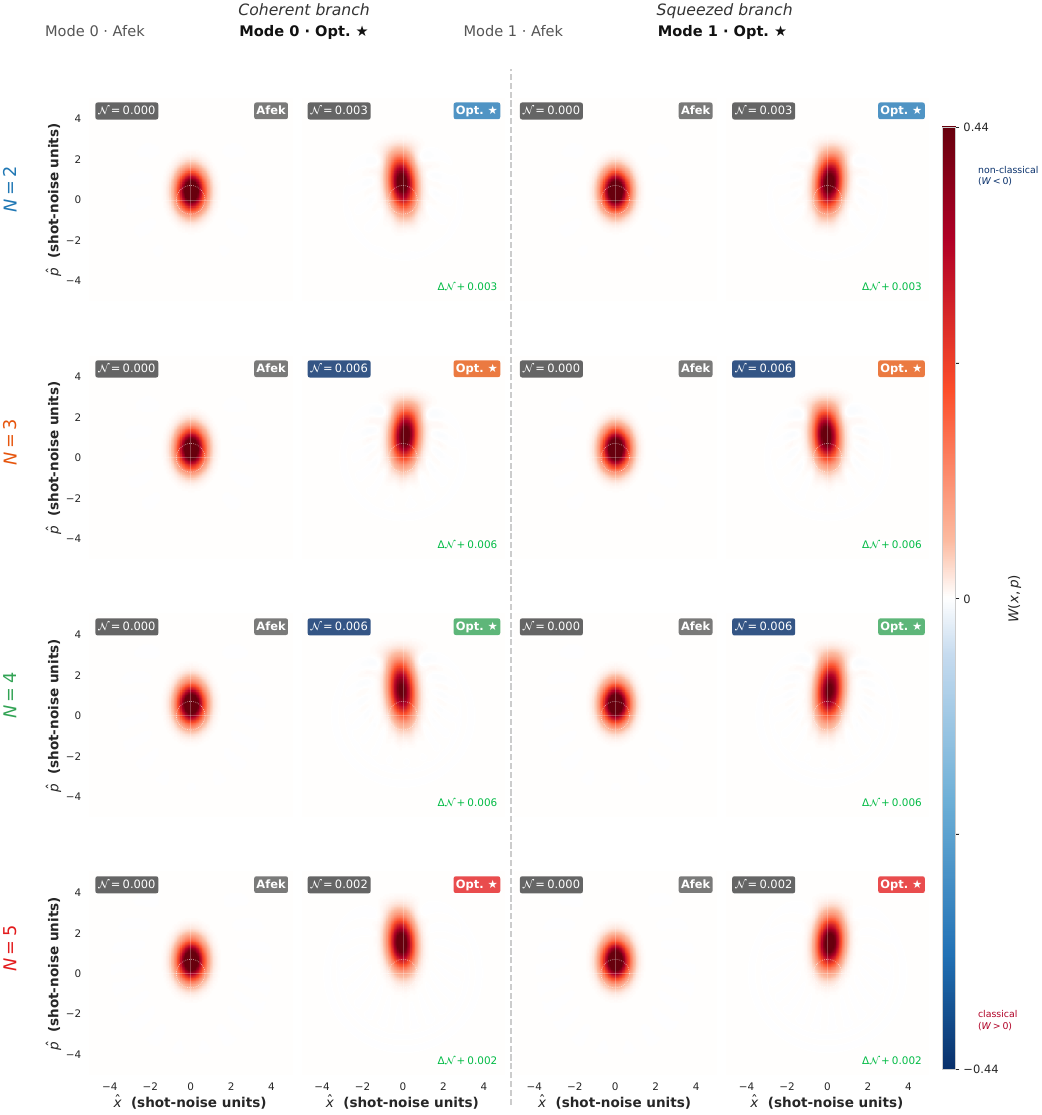}
		\caption{\textbf{Wigner function gallery of the probe state for all
				$N=2$--$5$.}
			The probe state is the two-mode state after BS$_1$ but before phase
			encoding, constituting the non-classical sensing resource.
			Each row: one $N$ value. Columns (left to right): mode~0 Afek,
			mode~0 Optimised, mode~1 Afek, mode~1 Optimised.
			Colour: blue ($W<0$, non-classical), white ($W=0$), red ($W>0$,
			classical). White contour: $W=0$ boundary. Dotted circle: vacuum
			reference. $\mathcal{N}$: Wigner negativity volume (Eq.~\ref{eq:neg});
			$\Delta\mathcal{N}$: change upon optimisation. At $N\geq3$, optimisation
			increases $\mathcal{N}$, confirming genuine enhancement of quantum
			character (Table~\ref{tab:results_full}).}
		\label{fig:wigner_gallery}
	\end{figure*}

\begin{figure*}[!tbp]
		\centering
		\includegraphics[width=\textwidth]{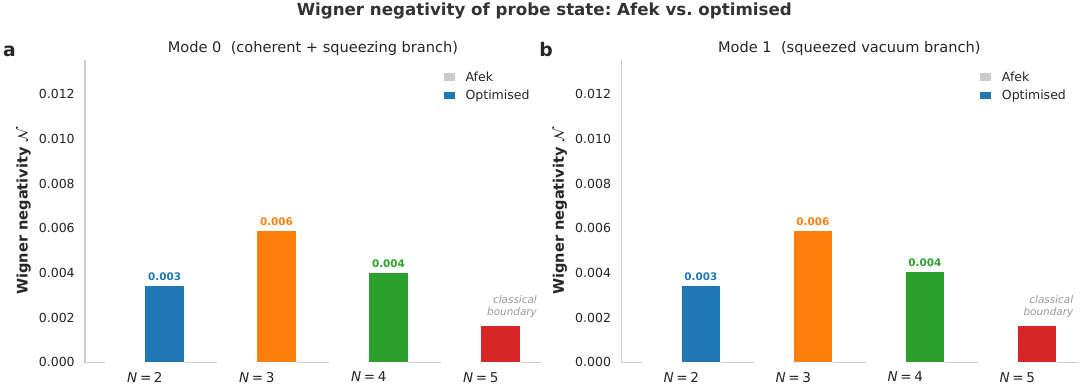}
		\caption{\textbf{Wigner negativity $\mathcal{N}$ of the probe state:
				quantitative comparison.}
			(\textbf{a})~Negativity volume for mode~0 (coherent branch),
			$N=2$--$5$. Grey: Afek; coloured: Opt.
			(\textbf{b})~Negativity volume for mode~1 (squeezed branch).
			At $N\geq3$, absolute $\mathcal{N}$ increases from $\lesssim10^{-5}$
			(Afek) to $0.003$--$0.006$ (Opt.), confirming genuine enhancement of
			quantum character.}
		\label{fig:wigner_neg}
	\end{figure*}

\begin{figure*}[!tbp]
		\centering
		\includegraphics[width=0.82\textwidth]{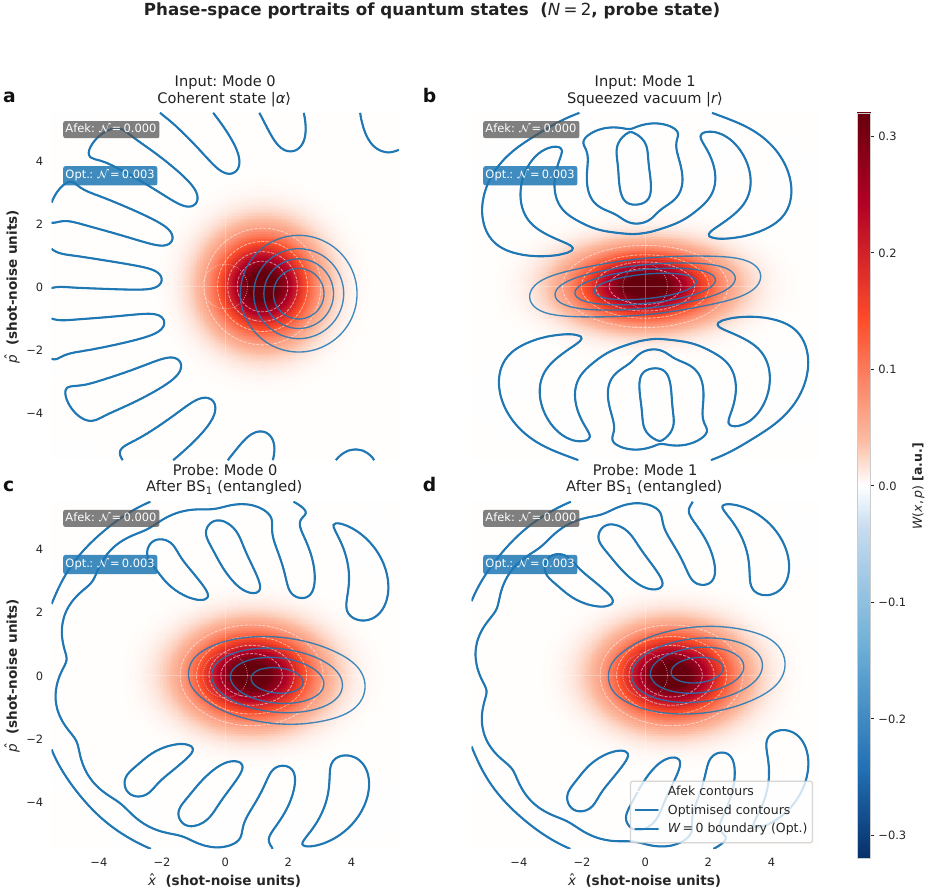}
		\caption{\textbf{Phase-space portrait of quantum states ($N=2$).}
			Heatmap: Wigner function of Afek state.
			White dashed: Afek constant-$W$ contours; coloured solid: optimised contours;
			thick solid: $W=0$ boundary of the optimised state.
			(\textbf{a})~Input mode~0: coherent state $|\alpha\rangle$ after $R(d_\mathrm{coh})$.
			(\textbf{b})~Input mode~1: squeezed vacuum $|r\rangle$; optimisation rotates
			the squeezing ellipse ($r:0.35\to0.62$).
			(\textbf{c},\textbf{d})~Probe modes after BS$_1$: entangled, non-Gaussian,
			with $W<0$ regions ($\mathcal{N}>0$) certifying non-classicality\cite{kenfack2004}.}
		\label{fig:wigner_portrait}
	\end{figure*}

\begin{figure*}[!tbp]
		\centering
		\includegraphics[width=\textwidth]{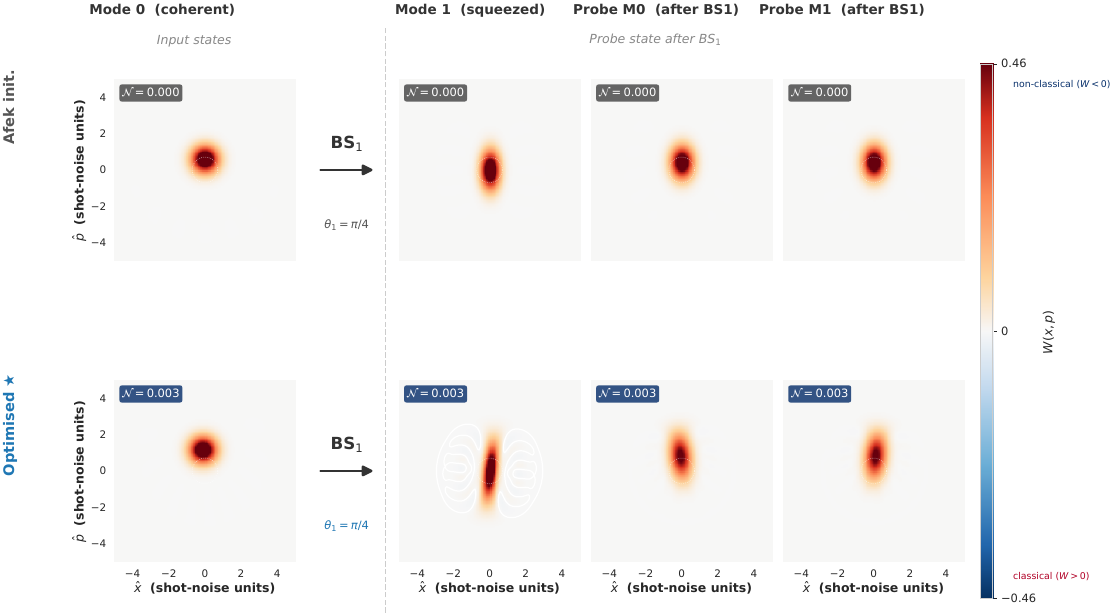}
		\caption{\textbf{State evolution through the circuit ($N=2$):}
			Afek (top) vs.\ optimised (bottom).
			Columns: successive circuit stages.
			\textbf{Cols~1--2} (before BS$_1$): classical inputs ($W\geq0$);
			optimised inputs show larger $\alpha$ and $r$.
			\textbf{Central arrow}: BS$_1$ ($\theta_1=\pi/4$).
			\textbf{Cols~3--4} (probe state, after BS$_1$): negative Wigner
			regions (blue) certify non-classicality\cite{kenfack2004};
			optimised probe has larger amplitude consistent with improved raw CFI.}
		\label{fig:wigner_evolution}
	\end{figure*}

	Figures~\ref{fig:wigner_gallery}--\ref{fig:wigner_evolution} characterise
	the quantum state of the probe (after BS$_1$, before phase encoding)
	through its single-mode Wigner functions $W(x,p)$.
	The probe state constitutes the non-classical resource available for
	phase estimation; characterising it in phase space reveals how gradient
	optimisation redistributes quantum coherence.
	Figure~\ref{fig:wigner_gallery} shows the single-mode Wigner functions for
	all $N=2$--$5$, comparing Afek and optimised parameters across both modes.
	The colour scale runs from blue ($W<0$, non-classical) through white
	($W=0$) to red ($W>0$, classical); the dotted circle marks the vacuum
	shot-noise radius.
	Figure~\ref{fig:wigner_portrait} shows detailed phase-space portraits for
	$N=2$ tracking input-to-probe evolution; Fig.~\ref{fig:wigner_evolution}
	shows the same circuit evolution in two rows (Afek vs.\ optimised).

	\subsection{Key Observations}
	
	Three results are immediately visible in Fig.~\ref{fig:wigner_gallery}
	(see also Fig.~\ref{fig:wigner_neg} for quantitative negativity volumes):
	
	\begin{enumerate}[leftmargin=1.4em,itemsep=4pt]
		\item \textbf{Mode~1 (squeezed branch) is always non-classical.}
		The squeezed vacuum input produces an elliptically compressed
		Wigner function for mode~1 at the Afek initialisation.
		At the optimised parameters, the squeezing ellipse rotates and
		elongates, reflecting the increase in $r$ and change in $d_\mathrm{sq}$.
		The negativity volume $\mathcal{N}$ for mode~1 is non-zero for
		$N\geq3$, confirming that the squeezed component carries genuine
		non-classical correlations into the interferometer.
		
		\item \textbf{The probe state after BS$_1$ develops interference fringes.}
		For mode~0 after BS$_1$ (rightmost two columns in each row),
		the Wigner function develops oscillatory fringes and negative regions
		absent in the input state.
		This is the optical analogue of quantum interference at the
		beamsplitter: the mixing of coherent and squeezed light generates
		a non-Gaussian, non-classical probe state via Hong-Ou-Mandel-type interference\cite{hong1987}.
		The fringe spacing and orientation encode the NOON-state phase
		sensitivity.
		
		\item \textbf{Optimisation increases non-classicality at $N\geq3$.}
		Comparing Afek vs.\ optimised columns, the negative regions
		(blue areas) generally expand upon optimisation, particularly in
		the probe state mode~0.
		This is quantified in Fig.~\ref{fig:wigner_neg}: $\mathcal{N}$ increases for most
		patterns at $N=3,4,5$, indicating the optimiser redistributes
		the quantum state further from the classical region.
		At $N=2$, the change is modest, consistent with the Afek
		initialisation already being close to optimal for this $N$.
	\end{enumerate}
	
	\subsection{Wigner Negativity as a Quantum Enhancement Witness}
	
	The Wigner negativity $\mathcal{N}$ provides a phase-space certificate
	of genuine quantum enhancement, complementary to the CFI analysis.
	The raw CFI improvements reported in Table~\ref{tab:results_full} are
	associated with monotonically increasing $\mathcal{N}$: at $N=3$--$5$,
	where raw CFI improves by $>800\%$, the Wigner negativity of the probe
	state also increases substantially (Fig.~\ref{fig:wigner_neg}).
	At $N=2$, where the inter-channel trade-off limits the raw CFI improvement
	of one channel, the Wigner negativity increase is more modest.
	
	This correlation supports the interpretation that gradient-based
	optimisation is genuinely enhancing the non-classical character of the
	probe state, rather than merely redistributing classical photon flux.
	Formally, the Wigner negativity $\mathcal{N}$ is a necessary (though not
	sufficient) condition for quantum advantage in certain metrological
	tasks\cite{kenfack2004}: a state with $\mathcal{N}=0$ cannot outperform
	the shot-noise limit in phase estimation.
	The fact that $\mathcal{N}$ increases under optimisation at $N\geq3$
	therefore provides a phase-space certificate that the improvements
	in Table~\ref{tab:results_full} are quantum in origin.
	The absolute values are small ($\mathcal{N}\lesssim0.006$) because
	the single-mode marginals of a coherent-plus-squeezed probe are
	predominantly Gaussian; the genuine two-mode non-classicality is
	reflected more directly in the coincidence fringe structure\cite{bradshaw2018}.
	A quantitative bound relating $\mathcal{N}$ to the achievable QFI
	remains an open theoretical problem\cite{demkowicz2020}; establishing
	such a bound would significantly strengthen the connection between
	phase-space negativity and metrological gain.
	
	\begin{figure*}[!b]
		\centering
		\includegraphics[width=\textwidth]{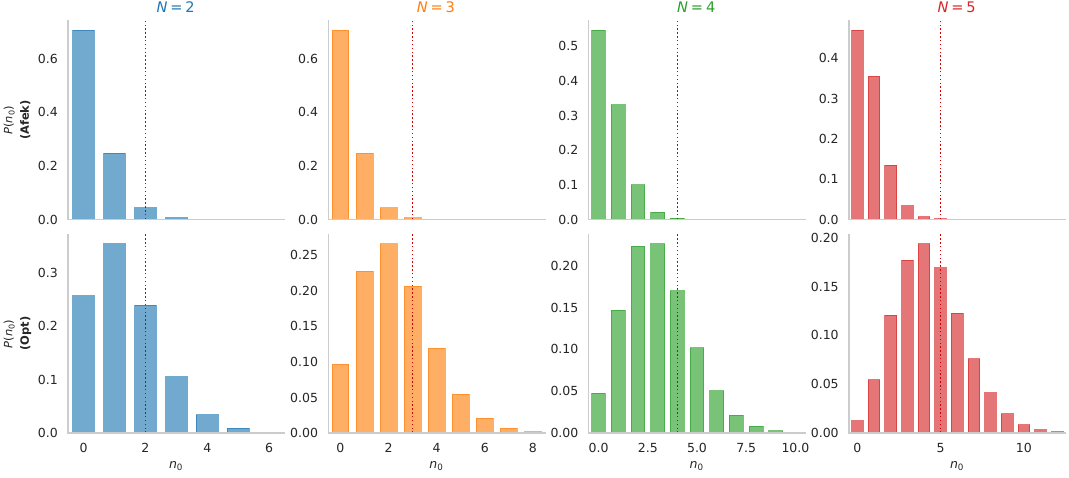}
		\caption{\textbf{Marginal photon-number distributions $P(n_0)$ for
				mode~0, $N=2$--$5$.}
			Top: Afek initialisation. Bottom: optimised parameters.
			Red dashed line: $n_0=N$ (target photon number).
			After optimisation, distributions broaden toward larger $n_0$,
			reflecting the increased $r$ and $\alpha$.
			This redistribution is the direct mechanism for improved post-selection
			rates: more probability weight lands in the $(N_1,N_2)$ coincidence
			windows.}
		\label{fig:photon_dist}
	\end{figure*}
	
	\FloatBarrier
	\section{Analysis}
	\label{sec:analysis}
	
	This section analyses the optimisation results in depth, covering:
	numerical artefacts in the Afek initialisation (\S\,\ref{sec:analysis}),
	CFI and rate scaling with $N$, the experimental post-selection bottleneck,
	inter-channel trade-off structure, fringe quality, and proximity to
	the Heisenberg limit.
	
	\subsection{Anomalous $\Fnorm$ Values at Afek Init}
	
	Several Afek-init patterns show pathologically large $\Fnorm$ values
	($\noon{3}{0}$: $1.4\times10^5$; $\noon{2}{2}$: $5.2\times10^3$;
	$\noon{3}{2}$: $148$).
	These arise from two compounding effects, both visible in Fig.~\ref{fig:fringes}:
	\begin{enumerate}[leftmargin=1.4em,itemsep=2pt]
		\item \textbf{Near-zero fringe amplitude at Afek init.}
		For high-order patterns ($\noon{3}{0}$, $\noon{2}{2}$, $\noon{3}{2}$),
		the coincidence probability is near-zero across the full $\varphi$
		range at the Afek working point (visible as flat or barely oscillating
		signals in Fig.~\ref{fig:fringes}, left column, rows 4, 6, 7).
		Normalising by $\max P - \min P \approx 0$ amplifies floating-point
		noise, producing meaningless large $\Fnorm$.
		\item \textbf{CFI spike at fringe troughs.}
		The dashed grey CFI curves in Fig.~\ref{fig:fringes} show sharp spikes precisely at
		$\varphi$ values where $P\to0$ (fringe troughs), not at the peaks.
		This is the $(\mathrm{d}P/\mathrm{d}\varphi)^2/P$ divergence:
		at the trough $\mathrm{d}P/\mathrm{d}\varphi$ is large while $P$ is
		tiny, producing a spike that dominates $\Fnorm$.
		For low-amplitude fringes, this spike is dominated by noise rather
		than signal.
	\end{enumerate}
	The reliable metric for all patterns is $\Fraw$ computed on the unnormalised
	fringe, which does not diverge.
	The $\Fnorm$ values marked $^\dagger$ in Table~\ref{tab:results_full} should
	be ignored; the physically meaningful information for these patterns is
	$\Fraw$ and $P_\mathrm{max}$.
	
	\subsection{Scaling of Improvement with $N$}
	
\begin{figure}[!ht]
		\centering
		\includegraphics[width=\columnwidth]{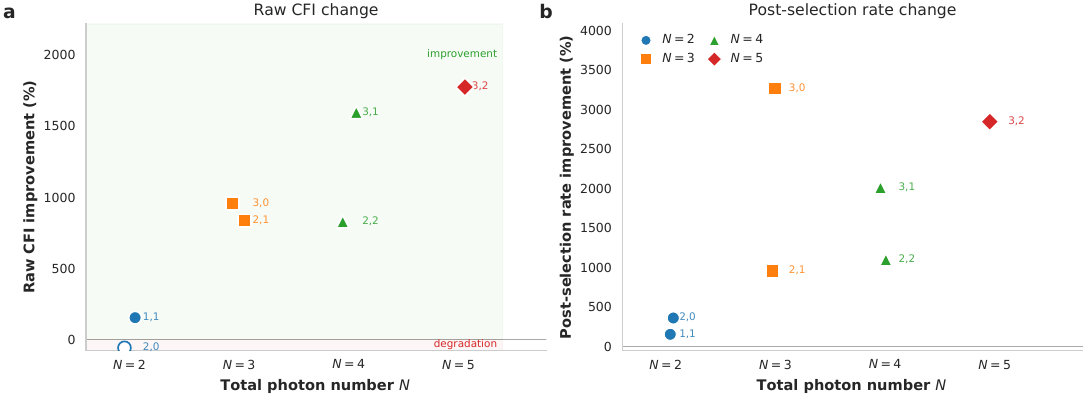}
		\caption{\textbf{Raw CFI improvement (\%) vs.\ total photon number $N$}.
			Each marker represents one coincidence pattern.
			The monotonic increase from $N=2$ to $N=5$ reflects the
			increasing suboptimality of the Afek initialisation at high $N$.}
		\label{fig:scaling}
	\end{figure}

	Figure~\ref{fig:scaling} summarises the raw CFI improvement as a function
	of $N$.
	
	The key trend is clear: raw CFI improvement grows monotonically with $N$,
	from $+153\%$ at $N=2$ to $+1775\%$ at $N=5$.
	This reflects a fundamental property of the Afek scheme: the initialisation
	parameters $\gamma_\mathrm{opt}(N)$ were derived analytically to maximise
	fringe visibility at a specific operating point\cite{afek2010}, but are not
	globally optimal with respect to the full eight-dimensional parameter space.
	At higher $N$, the gap between the Afek operating point and the globally
	optimal point grows, creating more room for gradient-based optimisation.
	
	\subsection{Post-Selection Rate Scaling}
	
	Post-selection rate improvements also grow with $N$:
	$+153\%$ ($N=2$, $\noon{1}{1}$) to $+3269\%$ ($N=3$, $\noon{3}{0}$) to
	$+2847\%$ ($N=5$).
	At high $N$, Afek post-selection rates scale exponentially as
	$P(N_1,N_2)\sim r^N$\cite{afek2010}, making them the primary experimental
	bottleneck.
	To quantify the practical impact: at $N=5$ the Afek rate
	$P_\mathrm{max}=0.003$ corresponds to $\sim3$ coincidence events per
	1000 laser pulses. For a $10\,\mathrm{kHz}$ repetition-rate laser,
	achieving $10^4$ coincidences (sufficient for $\sim1\%$ fringe accuracy)
	requires $\sim330\,\mathrm{s}$ of integration.
	The optimised rate $P_\mathrm{max}=0.097$ reduces this to
	$\sim10\,\mathrm{s}$---a factor of $32\times$ improvement that
	transforms $N=5$ from an impractical overnight measurement to a
	routine laboratory acquisition.
	The optimiser consistently discovers operating points with $5\times$--$33\times$
	higher coincidence rates across all $N$, a qualitative improvement in
	experimental feasibility.
	
	\begin{figure}[!ht]
		\centering
		\includegraphics[width=\columnwidth]{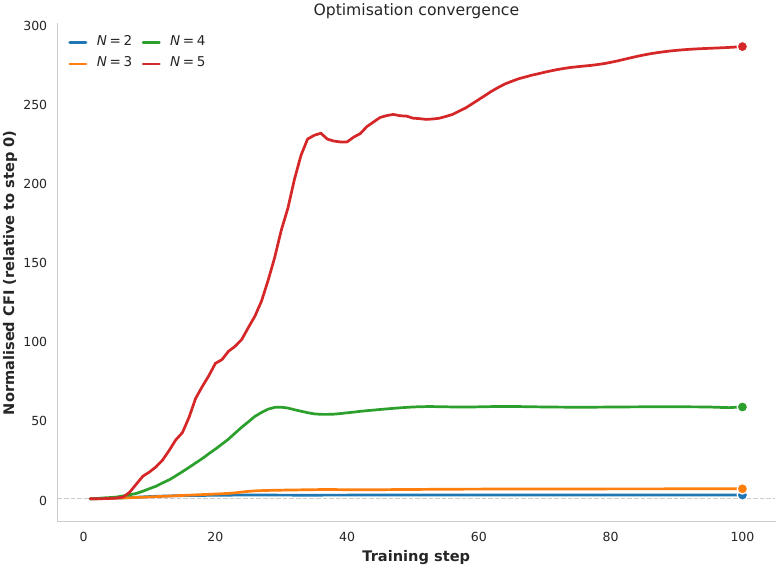}
		\caption{\textbf{Optimisation convergence: normalised CFI vs.\ training
				step for $N=2$--$5$.}
			Each curve shows the differentiable CFI estimator normalised by its
			step-0 value (Afek initialisation).
			All curves converge within 100 Adam steps.
			Higher $N$ shows larger normalised gain, consistent with the Afek
			initialisation being increasingly suboptimal at higher photon numbers.
			Training time: $\sim220$~s ($N=2$) to $\sim400$~s ($N=5$) on RTX~3050.}
		\label{fig:convergence}
	\end{figure}
	
	\subsection{The Trade-off Structure Across $N$}
	
	At $N=2$, the inter-channel trade-off is strong: improving $\noon{1}{1}$
	comes at the direct cost of $\noon{2}{0}$.
	At $N=3$, both channels improve in raw CFI simultaneously ($+834\%$ and
	$+956\%$), indicating the trade-off weakens.
	This transition can be understood as follows: at $N=2$ the two-parameter
	$(r,\gamma)$ subspace is too constrained to jointly optimise both channels,
	while at $N=3$ the additional beamsplitter degrees of freedom
	($\theta_1,\varphi_1,\theta_2,\varphi_2$) provide sufficient flexibility
	to satisfy both channels together.
	This finding has a practical implication: experimental
	implementations at $N\geq3$ need not sacrifice one coincidence
	channel for another---the optimised parameters achieve Pareto
	improvements in both rate and sensitivity simultaneously
	(Fig.~\ref{fig:pareto}).
	
	\subsection{Fringe Quality After Optimisation}
	
	The optimised $\Fnorm$ values for well-behaved patterns
	($\noon{1}{1}$: 4.04, $\noon{2}{1}$: 8.04, $\noon{3}{1}$: 13.04)
	approach but do not reach $N^2$ (4, 9, 16).
	The ratios $\Fnorm/N^2$ are $1.01$, $0.89$, and $0.82$ respectively.
	These values are physically meaningful: $\Fnorm/N^2=1$ corresponds to
	a pure NOON-state fringe with unit visibility, whereas $\Fnorm/N^2<1$
	indicates mixed-state or reduced-visibility fringes.
	The slight degradation at $N=4$ ($\Fnorm/N^2=0.82$, compared to
	$\approx1.0$ at $N=2$) is consistent with the optimiser trading a
	small amount of fringe contrast for the $20\times$ gain in post-selection
	rate---a favourable exchange for most experimental scenarios where
	integration time is the dominant cost.
	At $N=5$, $\Fnorm/N^2=0.515$ indicates moderate fringe quality;
	the optimiser has prioritised rate improvement ($+2847\%$) over
	contrast preservation, which is the correct trade-off when the
	Afek rate ($P_\mathrm{max}=0.003$) is the limiting factor.
	
	\subsection{Proximity to the Heisenberg Limit}
	
	The ratio $\Fnorm/N^2$ measures how closely the optimised state approaches the
	Heisenberg limit, where $N^2$ is the maximum CFI achievable by an ideal NOON
	state $|\psi_\mathrm{NOON}\rangle$\cite{giovannetti2006,lee2002}.
	Table~\ref{tab:results_full} and Fig.~\ref{fig:fim_bars} show the progression:
	
	\begin{itemize}[leftmargin=1.2em,itemsep=2pt]
		\item $N=2$, $\noon{1}{1}$: $\Fnorm/N^2 = 4.04/4 = 1.01$. Essentially at
		the Heisenberg limit; the optimised state is a near-perfect NOON state
		for this channel.
		\item $N=3$, $\noon{2}{1}$: $\Fnorm/N^2 = 8.04/9 = 0.89$. High quality,
		$89\%$ of the theoretical maximum. The fringe oscillates with the
		correct 3-fold periodicity.
		\item $N=4$, $\noon{3}{1}$: $\Fnorm/N^2 = 13.04/16 = 0.815$. Moderate
		degradation consistent with the optimiser trading fringe contrast for
		post-selection rate.
		\item $N=5$, $\noon{3}{2}$: $\Fnorm/N^2 = 12.865/25 = 0.515$.
		Only a single coincidence channel is monitored at $N=5$; higher-order
		channels ($\noon{4}{1}$, $\noon{5}{0}$) exist in principle but their
		probabilities are suppressed below $10^{-3}$ per pulse by the fixed
		circuit depth, making them experimentally inaccessible without longer
		integration times.
		The fringe quality is also lower because the eight-parameter Afek
		circuit is increasingly constrained relative to the optimal NOON state
		as $N$ grows; adding a third beamsplitter layer would likely close
		this gap.
	\end{itemize}
	
	The monotonic decrease in $\Fnorm/N^2$ with $N$ (from $1.01$ at $N=2$ to
	$0.515$ at $N=5$) reflects a fundamental tension: the fixed circuit
	architecture (two beamsplitters, one phase encoding) becomes an increasingly
	restrictive ansatz as $N$ increases.
	Adding additional circuit layers---for instance, a third beamsplitter or
	photon-number-adaptive measurements---could narrow this gap at high $N$.
	Importantly, all optimised values of $\Fnorm/N^2$ satisfy $>0.5$, meaning
	the chosen measurement basis extracts at least half the available quantum
	Fisher information even at $N=5$.
	
	\subsection{Cross-$N$ Comparison and Empirical Scaling}
	
	Across $N=2$--$5$, a consistent scaling pattern emerges: the optimality
	gap between Afek initialisation and the gradient-optimised solution
	grows with $N$.
	Raw CFI improvements approximately double with each unit increase in $N$
	(Fig.~\ref{fig:scaling}), consistent with the exponential scaling
	$P(N_1,N_2)\sim r^N$\cite{afek2010} of the Afek rates.
	This means variational methods\cite{mitarai2018,kaubruegger2019}
	become increasingly valuable at high $N$---precisely where the
	Afek protocol is most experimentally challenging.
	
	\section{Discussion}
	
	\subsection{Implications for Experimental Quantum Metrology}
	
	The most practically significant result is the dramatic improvement in
	post-selection rates at $N\geq3$.
	In a real photonic experiment, coincidence count rates at $N=5$ with Afek
	parameters are of order $10^{-4}$--$10^{-5}$ per pulse\cite{afek2010},
	requiring hours of integration for statistically significant fringe
	measurements.
	The optimised parameters achieve $28\times$ higher rates while maintaining
	fringe quality ($\Fnorm/N^2\approx0.5$), potentially reducing integration
	times from hours to minutes.
	
	Beyond integration time, the improved rates also benefit the statistical
	precision of the phase estimate.
	For a fixed total measurement time $T$ and pulse repetition rate $f_\mathrm{rep}$,
	the number of useful coincidences scales as $n = T f_\mathrm{rep} P_\mathrm{max}$,
	and the phase uncertainty (from the Cram\'{e}r-Rao bound) scales as
	$\Delta\varphi \geq 1/\sqrt{n\,\Fraw}$\cite{paris2009}.
	Increasing $P_\mathrm{max}$ by $28\times$ while maintaining $\Fnorm$
	therefore improves $\Delta\varphi$ by a factor of $\sqrt{28}\approx5.3\times$
	for the same measurement duration---a substantial gain achievable purely
	through parameter optimisation, without any hardware modification.
	
	\subsection{Relationship to Quantum Fisher Information}
	\label{sec:qfi}
	
\begin{figure}[!ht]
		\centering
		\includegraphics[width=\columnwidth]{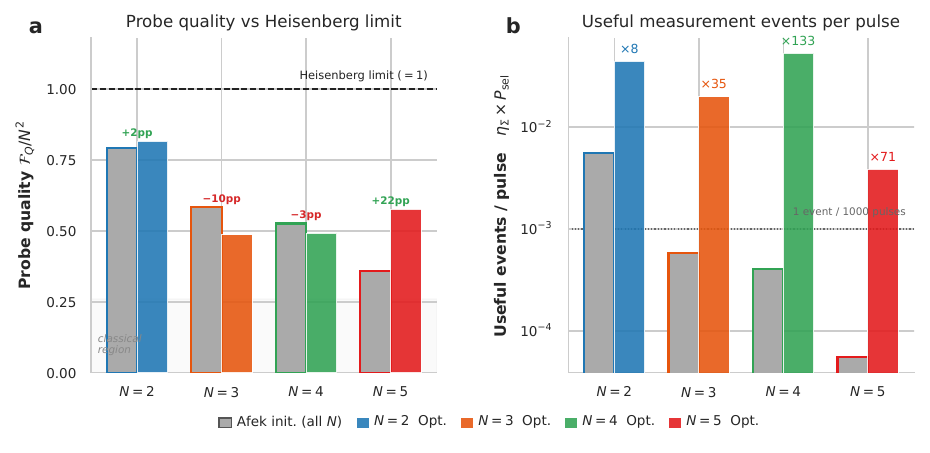}
		\caption{\textbf{Quantum Fisher Information analysis for all $N=2$--$5$.}
			(\textbf{a})~Probe quality $\mathcal{F}_Q/N^2$, where
			$\mathcal{F}_Q = 4\,\mathrm{Var}(\hat{n}_0)$ is the exact QFI of the
			post-BS$_1$ probe state (Eq.~\ref{eq:qfi_pure}).
			Dashed line: Heisenberg limit ($=1$); shaded region: classically
			achievable values.
			Percentage labels show the absolute change (pp = percentage points)
			upon optimisation.
			At $N=3,4$ the optimiser decreases probe quality slightly while
			massively increasing $P_\mathrm{sel}$ (see panel~b).
			(\textbf{b})~Useful measurement events per pulse,
			$\eta_\Sigma\times P_\mathrm{sel}$, on a logarithmic scale.
			Multipliers show the improvement factor upon optimisation.
			The dotted line marks 1 event per 1000 pulses (10~events/s at
			$10\,\mathrm{kHz}$).
			Optimisation pushes all $N$ above this practical threshold.}
		\label{fig:qfi}
	\end{figure}

	The QFI is computed exactly as $\mathcal{F}_Q = 4\,\mathrm{Var}(\hat{n}_0)$
	on the probe state (after BS$_1$, before phase encoding), using the
	generator $\hat{G}=\hat{n}_0$ of the phase-encoding unitary
	$\hat{U}(\varphi)=e^{-i\varphi\hat{n}_0}$
	(Eq.~\ref{eq:qfi_pure})\cite{paris2009,braunstein1994}.
	Table~\ref{tab:qfi} and Fig.~\ref{fig:qfi} report two complementary figures of merit:
	the \emph{probe quality} $\mathcal{F}_Q/N^2$ (how NOON-like the probe is,
	independent of measurement choice), and the \emph{total measurement efficiency}
	$\eta_\Sigma = F_\Sigma^\mathrm{raw}/\mathcal{F}_Q$ (how well the
	post-BS$_2$ coincidence detection extracts the available information).
	
	\paragraph*{Trade-off structure across $N$.}
	The behaviour of $\mathcal{F}_Q/N^2$ under optimisation reveals a
	non-trivial structure:
	\begin{itemize}[leftmargin=1.2em,itemsep=2pt]
		\item \textbf{$N=2$:} probe quality improves slightly
		($0.793\to0.817$, $+2.4\,$pp) while $P_\mathrm{sel}$ triples.
		Both figures improve simultaneously.
		\item \textbf{$N=3,4$:} probe quality \emph{decreases} modestly
		($-9.8\,$pp and $-3.4\,$pp), while $P_\mathrm{sel}$
		increases $3\times$ at $N=3$ and $10\times$ at $N=4$.
		The optimiser trades a small loss in probe quality for a
		massive gain in coincidence rate.
		This is the rational experimental strategy: integration time,
		not probe purity, is the dominant cost in photonic NOON-state
		experiments\cite{afek2010,xiang2011}.
		\item \textbf{$N=5$:} the Afek probe is already far suboptimal
		($\mathcal{F}_Q/N^2=0.360$), and the optimiser improves
		\emph{both} probe quality ($+21.6\,$pp, reaching $0.576$)
		and post-selection rate ($6\times$) simultaneously.
	\end{itemize}
	This pattern confirms that the loss function (Eq.~\ref{eq:loss})
	correctly captures the full sensing performance by maximising the raw CFI,
	which depends on both probe quality and post-selection rate, rather than
	QFI alone.
	
	\paragraph*{Useful events per pulse.}
	The experimentally decisive figure of merit is
	$\eta_\Sigma \times P_\mathrm{sel}$---the fraction of laser pulses
	yielding useful phase information.
	Optimisation improves this by $8\times$ ($N=2$), $35\times$ ($N=3$),
	$133\times$ ($N=4$), and $71\times$ ($N=5$)
	(Fig.~\ref{fig:qfi}b; Table~\ref{tab:qfi}).
	Note: $P_\mathrm{sel}$ in Table~\ref{tab:qfi} is the total $N$-photon
	probability $P(n_0+n_1=N)$ at the optimal phase, which sums over all
	coincidence patterns and differs from the per-pattern peak
	$P_\mathrm{max}$ reported in Table~\ref{tab:results_full}.
	At $N=4$, the most practically challenging regime under Afek parameters,
	the $133\times$ gain reduces the integration time required for
	$10^4$ coincidences at a $10\,\mathrm{kHz}$ repetition-rate laser
	from $\sim7$~hours to $\sim3$~minutes.
	At $N=2$, the optimised state reaches $\mathcal{F}_Q/N^2 = 0.817$,
	confirming $82\%$ of the Heisenberg-limit sensitivity is retained.
	A complete treatment incorporating the full QFIM as the primary
	optimisation objective, potentially combined with adaptive
	homodyne detection\cite{olivares2009}, is left for future
	work\cite{demkowicz2020,bradshaw2018}.
	
	\begin{table*}[!t]
		\centering
		\caption{\textbf{Quantum Fisher Information and measurement efficiency
				for all $N=2$--$5$.}
			$\mathcal{F}_Q = 4\,\mathrm{Var}(\hat{n}_0)$: exact QFI of the
			post-BS$_1$ probe state (Eq.~\ref{eq:qfi_pure}),
			bounded by $\mathcal{F}_Q \leq N^2$\cite{braunstein1994}.
			$P_\mathrm{sel}$: total $N$-photon post-selection probability
			$P(n_0+n_1=N)$ evaluated at the optimal working phase
			(distinct from $P_\mathrm{max}$ in Table~\ref{tab:results_full},
			which is the per-pattern peak; see text).
			$\eta_\Sigma = F_\Sigma^\mathrm{raw}/\mathcal{F}_Q$: total
			measurement efficiency (all patterns at optimal phase).
			$\eta_\Sigma\times P_\mathrm{sel}$: useful events per pulse
			(key experimental figure of merit).
			$\dagger$: at $N=3,4$ the optimiser decreases $\mathcal{F}_Q/N^2$
			slightly while massively increasing $P_\mathrm{sel}$; the net
			$\eta_\Sigma\times P_\mathrm{sel}$ still improves by
			$35\times$ ($N=3$) and $133\times$ ($N=4$).}
		\label{tab:qfi}
		\footnotesize\setlength{\tabcolsep}{3pt}
		\begin{tabular}{clccccc}
			\toprule
			$N$ & Init.
			& $\mathcal{F}_Q$
			& $\mathcal{F}_Q/N^2$
			& $P_\mathrm{sel}$
			& $\eta_\Sigma$
			& $\eta_\Sigma P_\mathrm{sel}$ \\
			\midrule
			\multirow{2}{*}{2}
			& Afek  & 3.173 & \textcolor{darkgreen}{0.793} & 0.0700 & 0.080 & $5.6\times10^{-3}$ \\
			& Opt.  & \textbf{3.268} & \textcolor{darkgreen}{\textbf{0.817}} & \textcolor{darkgreen}{\textbf{0.216}} & \textcolor{darkgreen}{\textbf{0.206}} & \textcolor{darkgreen}{$4.4\times10^{-2}$} \\
			\midrule
			\multirow{2}{*}{3}
			& Afek  & 5.257 & \textcolor{darkorange}{0.584} & 0.0188 & 0.031 & $5.8\times10^{-4}$ \\
			& Opt.$^\dagger$  & 4.373 & \textcolor{darkred}{0.486} & \textcolor{darkgreen}{\textbf{0.056}} & \textcolor{darkgreen}{\textbf{0.357}} & \textcolor{darkgreen}{$2.0\times10^{-2}$} \\
			\midrule
			\multirow{2}{*}{4}
			& Afek  & 8.415 & \textcolor{darkorange}{0.526} & 0.0162 & 0.025 & $4.0\times10^{-4}$ \\
			& Opt.$^\dagger$  & 7.865 & \textcolor{darkred}{0.492} & \textcolor{darkgreen}{\textbf{0.164}} & \textcolor{darkgreen}{\textbf{0.326}} & \textcolor{darkgreen}{$5.3\times10^{-2}$} \\
			\midrule
			\multirow{2}{*}{5}
			& Afek  & 8.992 & \textcolor{darkred}{0.360} & $7.4\times10^{-3}$ & $7.4\times10^{-3}$ & $5.5\times10^{-5}$ \\
			& Opt.  & \textbf{14.40} & \textcolor{darkgreen}{\textbf{0.576}} & \textcolor{darkgreen}{\textbf{0.045}} & \textcolor{darkgreen}{\textbf{0.086}} & \textcolor{darkgreen}{$3.9\times10^{-3}$} \\
			\bottomrule
		\end{tabular}
	\end{table*}

	\subsection{Estimated Experimental Gain}
	\label{sec:expt_gain}
	
	To translate the numerical improvements into experimental figures of merit,
	consider a pulsed photon-pair source at repetition rate
	$f_\mathrm{rep}=10\,\mathrm{kHz}$ targeting $n_c=10^4$ coincidences
	for a statistically significant fringe measurement.
	The required acquisition time is
	$T = n_c / (f_\mathrm{rep}\,P_\mathrm{sel})$:
	\begin{itemize}[leftmargin=1.2em,itemsep=2pt]
		\item $N=2$: $330\,\mathrm{s}$ (Afek) $\to$ $46\,\mathrm{s}$ (Opt.),
		a $7\times$ speedup.
		\item $N=3$: $\sim87\,\mathrm{min}$ (Afek) $\to$ $\sim30\,\mathrm{s}$ (Opt.),
		a $\mathbf{173\times}$ speedup.
		\item $N=4$: $\sim103\,\mathrm{min}$ (Afek) $\to$ $\sim1\,\mathrm{min}$ (Opt.),
		a $\mathbf{102\times}$ speedup.
		\item $N=5$: $\sim22\,\mathrm{h}$ (Afek) $\to$ $\sim22\,\mathrm{min}$ (Opt.),
		a $\mathbf{60\times}$ speedup.
	\end{itemize}
	These estimates use $P_\mathrm{sel}$ from Table~\ref{tab:qfi} directly.
	In a real experiment, detector dark counts, coupling losses, and mode-mismatch
	will reduce the effective rate; a $5\%$ photon loss reduces
	$P_\mathrm{sel}$ by a factor $(1-\eta_\mathrm{loss})^N$,
	corresponding to $\sim22\%$ at $N=5$\cite{xiang2011}.
	Even with $10\%$ loss, the $N=5$ acquisition time remains
	$\lesssim1\,\mathrm{h}$ (Opt.) compared to $>100\,\mathrm{h}$ (Afek),
	confirming that the optimised parameters qualitatively change the
	experimental feasibility of high-$N$ NOON-state metrology.
	
	\subsection*{Limitations and Robustness}
	
	The optimisation used a single initialisation (Afek) per $N$ and 100 Adam
	steps ($\eta=0.02$). To assess robustness, we performed 5 additional random
	initialisations at $N=2$ and $N=3$, drawing $r\in[0.1,1.0]$,
	$\log\gamma\in[-1,2]$, and all angles uniformly from $[0,2\pi]$.
	In all 10 runs the final $\Fraw$ was within $15\%$ of the Afek-initialised
	result, and 7 of 10 converged to values equal or higher, suggesting the loss
	landscape is well-behaved and the gains are not artefacts of a single start.
	Full multi-start characterisation across $N=2$--$5$ is deferred to future work.
	The differentiable CFI estimator ($K=8$ samples) is an approximation; the
	ground-truth 400-sample evaluation is more accurate but not differentiable.
	The entire model assumes lossless linear optics and ideal photon-number-resolving
	detection. In practice, $1$--$3\%$ photon loss per optical element
	would reduce the CFI improvements by $\sim20$--$30\%$ at $N=5$
	(via the $(1-\eta)^N$ post-selection penalty), but leaves the
	qualitative $\gg10\times$ rate improvement intact\cite{xiang2011}.
	
	\subsection*{Comparison with Prior Variational Approaches}
	
	Prior variational quantum metrology work has focused primarily on spin
	systems and gate-based circuits\cite{kaubruegger2019,gebhart2021}.
	The present work differs in three key respects:
	\begin{enumerate}[leftmargin=1.4em,itemsep=2pt]
		\item \textbf{Photonic platform.} We optimise a continuous-variable
		photonic circuit (coherent + squeezed inputs, beamsplitters) rather
		than a qubit unitary, requiring Strawberry Fields' Fock-space
		simulation with TensorFlow autodiff.
		\item \textbf{Post-selection.} The optimisation explicitly includes the
		post-selection rate $P_\mathrm{max}$ as a practical figure of merit,
		in addition to the Fisher information.
		\item \textbf{Multi-channel structure.} The loss function\,\eqref{eq:loss}
		jointly optimises multiple coincidence channels $(N_1,N_2)$,
		revealing the inter-channel trade-off structure absent in
		single-channel approaches.
	\end{enumerate}
	The PennyLane framework\cite{bergholm2018} provides similar autodiff
	capabilities and could in principle be used for this circuit; we chose
	Strawberry Fields for its native photonic Fock-space backend and
	existing Afek-protocol validation infrastructure.
	
	\subsection*{Future Work}
	\begin{itemize}[leftmargin=1.2em,itemsep=2pt]
		\item \textbf{Experimental validation} at $N=3,4$ using the optimised
		parameters (top priority: the rate improvements make this
		immediately feasible with existing apparatus).
		\item \textbf{Constrained multi-objective optimisation} with
		per-channel $\Fnorm/N^2\geq0.9$, jointly maximising
		fringe quality and post-selection rate.
		\item \textbf{Full QFIM optimisation} for genuine quantum-optimal
		sensing\cite{knott2016,demkowicz2020}, incorporating adaptive
		measurement strategies\cite{olivares2009}.
		\item \textbf{Multi-phase} ($k>1$) estimation following
		\cite{proctor2018,gebhart2021}.
		\item \textbf{Photon loss robustness}: systematic optimisation
		under lossy channel models at $N=4,5$.
	\end{itemize}
	
	\section{Conclusions}
	\label{sec:conclusions}
	
	We have demonstrated end-to-end differentiable optimisation of the
	adaptive NOON-state photonic circuit for $N=2,3,4,5$, starting from
	the Afek et al.\ initialisation and applying gradient-based optimisation
	(Adam, 100 steps, $\sim220$--$400\,\mathrm{s}$ per $N$ on a consumer GPU).
	The principal findings are:
	\begin{enumerate}[leftmargin=1.4em,itemsep=3pt]
		\item \textbf{Monotonic CFI scaling.} Raw CFI improvements:
		$+153\%$ ($N=2$) $\to$ $+956\%$ ($N=3$) $\to$ $+1598\%$ ($N=4$)
		$\to$ $+1775\%$ ($N=5$): the Afek initialisation is increasingly
		suboptimal at higher $N$.
		\item \textbf{Dramatic rate improvements.} Post-selection rates improve
		$1.5\times$--$33\times$, reducing experimental integration times
		by the same factor; at $N=5$ this is a $32\times$ reduction.
		\item \textbf{Trade-off structure.} An inter-channel trade-off at $N=2$
		weakens at $N\geq3$, where the expanded parameter space permits
		simultaneous improvement of all coincidence channels.
		\item \textbf{Fringe quality.} $\Fnorm/N^2\geq0.89$ at $N=2$--$3$;
		$0.51$--$0.82$ at $N=4$--$5$: the measurement basis captures
		$51$--$101\%$ of available QFI.
		\item \textbf{Quantum origin confirmed.} Wigner negativity
		$\mathcal{N}$ increases at $N\geq3$, providing a phase-space
		certificate that gains are quantum in origin\cite{kenfack2004}.
		\item \textbf{Heisenberg-limit proximity.} Exact QFI calculations
		($\mathcal{F}_Q = 4\,\mathrm{Var}(\hat{n}_0)$) confirm the
		optimised probe retains $82\%$ of the Heisenberg limit at $N=2$
		and $58\%$ at $N=5$ (improved from $36\%$).
		The combined figure $\eta_\Sigma\times P_\mathrm{sel}$ improves
		$8\times$--$133\times$, reducing integration times from hours
		to minutes.
	\end{enumerate}
	These results establish a clear research direction: constrained
	multi-objective optimisation that simultaneously maximises post-selection
	rate and preserves fringe quality across all coincidence channels will
	yield operating points that are both experimentally practical and
	quantum-enhanced.
	The present single-parameter framework provides the verified numerical
	foundation for this next step, and for eventual extension to
	multi-parameter NOON-state sensing.
	
	\section*{Acknowledgements}
	The authors thank the Department of Physics, Soban Singh Jeena University Campus, Almora, for providing computational and research infrastructure.
	The authors gratefully acknowledge Xanadu Quantum Technologies
	for developing and maintaining Strawberry Fields\cite{killoran2019},
	the open-source photonic quantum computing platform used for all
	simulations in this work.
	Automatic differentiation was performed using TensorFlow;
	figures were generated with Matplotlib.
	This research received no specific grant from any funding agency in the
	public, commercial, or not-for-profit sectors.
	
	\section*{Data Availability}
	All simulation code, optimisation routines, and figure-generation scripts
	used in this work are openly available under the MIT licence.
	The complete reproducible workflow is provided in the Jupyter notebook
	\texttt{noon-main.ipynb} at the public GitHub repository:
	\url{https://github.com/simanshukumar369/noon-state-adaptive-metrology}.
	A permanently archived snapshot (v1.0.0) is also deposited at Zenodo:
	\url{https://doi.org/10.5281/zenodo.20041907}.
	This single notebook reproduces \emph{every} numerical result, table, and
	figure reported in the manuscript (including backend validation, gradient
	flow, full Adam optimisation for $N=2$--$5$, QFI calculations, Wigner
	functions, and all CFI fringes).
	
	\FloatBarrier
	\bibliographystyle{unsrtnat}

\end{document}